\documentclass[prd,onecolumn,nofootinbib,showpacs,showkeys]{revtex4}
\usepackage{bm,amsmath,amssymb}

\begin{document}

\title{Affine theory of gravitation}
\author{Nikodem Pop{\l}awski}
\email{nikodem.poplawski@gmail.com}
\date{\today}

\begin{abstract}
We propose a new theory of gravitation, in which the affine connection is the only dynamical variable describing the gravitational field.
We construct the simplest dynamical Lagrangian density that is entirely composed from the connection, via its curvature and torsion, and is an algebraic function of its derivatives.
It is given by the contraction of the Ricci tensor with a tensor which is inverse to the symmetric, contracted square of the torsion tensor, $k_{\mu\nu}=S^\rho_{\phantom{\rho}\lambda\mu}S^\lambda_{\phantom{\lambda}\rho\nu}$.
We vary the total action for the gravitational field and matter with respect to the affine connection, assuming that the matter fields couple to the connection only through $k_{\mu\nu}$.
We derive the resulting field equations and show that they are identical with the Einstein equations of general relativity with a nonzero cosmological constant, if the tensor $k_{\mu\nu}$ is regarded as the metric tensor.
The cosmological constant is simply a constant of proportionality between the two tensors, which together with $c$ and $G$ provides a natural system of units in gravitational physics.
This theory therefore provides a physically valid construction of the metric as an algebraic function of the connection, and naturally explains dark energy as an intrinsic property of spacetime.
The observed accelerating expansion of the Universe may be the strongest evidence for torsion.
\end{abstract}


\maketitle

\section{Introduction}

Einstein's general theory of relativity (GR) is based on the metric tensor as a fundamental variable describing a gravitational field \cite{LL}.
The affine connection, which defines a covariant differentiation in curved spacetime, is entirely composed from the metric.
Accordingly, the antisymmetric part of the connection (the torsion tensor) is assumed to vanish.
Although GR has been confirmed by many astrophysical observations, it predicts the formation of unphysical curvature singularities in black holes and at the big bang \cite{LL,OS}, and does not explain the observed acceleration of the Universe (dark energy) \cite{DE}.
The simplest form of dark energy is the cosmological constant, which can be added into the gravitational action by hand.
Such a term, however, has no explanation of its origin.

The problem of singularities may be resolved by abandoning the GR condition that the affine connection be symmetric and regarding the torsion tensor as a dynamical variable \cite{KS,Ham}.
If the gravitational Lagrangian density is proportional to the curvature scalar, like in GR, then the resulting theory is the simplest one with torsion and called the Einstein-Cartan-Sciama-Kibble (ECSK) theory of gravity \cite{KS,Hehl}.
In the presence of Dirac spinor fields, the torsion tensor is different from zero as a consequence of the Einstein-Cartan field equations, and is proportional to the Dirac spin density.
The coupling between spinors and torsion appears to avoid the initial cosmological singularity, replacing the big bang by a nonsingular big bounce \cite{avert,bb}.
It also explains why the present Universe appears spatially flat, homogeneous and isotropic at largest scales, without cosmic inflation \cite{bb,infl}.

In the ECSK theory of gravity, which is a first-order theory, two different geometrical objects (metric and connection) are dynamical variables.
The number of degrees of freedom of spacetime in this theory is thus bigger than that in GR.
It is possible, however, to further modify the structure of spacetime by constructing a Lagrangian density without the metric and defining the metric tensor as the derivative of the Lagrangian density with respect to the symmetrized Ricci tensor \cite{Sch}.
A simple Lagrangian density, found by Eddington, is given by the square root of the absolute value of the determinant of the symmetrized Ricci tensor \cite{Ed}.
Schr\"{o}dinger has shown that such a Lagrangian density leads, through the stationarity of the action with respect to variations of the (asymmetric) connection, to the Einstein field equations in vacuum with a cosmological constant \cite{Sch}.

That purely affine formulation of gravity explains why the metric Lagrangian for a gravitational field is linear in the Riemann curvature tensor \cite{FK}.
It reproduces the metric field equations for an electromagnetic field, described by the Maxwell Lagrangian density in which the metric tensor is replaced by the symmetrized Ricci tensor \cite{Max}.
However, adding the affine Eddington and Maxwell Lagrangian densities does not reproduce the metric field equations resulting from the sum of the metric Lagrangian densities; the affine field equations become more complicated \cite{Max_cc}.
Constructing purely affine Lagrangians for other matter fields also encounters difficulties \cite{fl}.

The problems of that formulation of gravity may be related to its definition of the metric tensor, which contains the total Lagrangian density.
Such a tensor is not a purely geometrical object constructed solely from the affine connection; its definition involves the matter fields.
Furthermore, the determinant and the inverse of the symmetrized Ricci tensor, which appear in the affine Lagrangians of \cite{Max,Max_cc,fl}, are not algebraic functions of derivatives of the connection.
In this paper, we propose a simpler dynamical Lagrangian density that is entirely composed from the affine connection and is a linear function of its derivatives.
It is given by the contraction of the Ricci tensor with the inverse of the symmetric, contracted square of the torsion tensor, multiplied by the determinant of the torsion square.
We show that the resulting field equations, if the matter fields couple to the connection only through the torsion square, are equivalent to the Einstein equations with a cosmological constant.

\section{Affine field Lagrangian}

Since the affine connection $\Gamma^{\,\,\rho}_{\mu\,\nu}$ is not a tensor, it can appear in an affine Lagrangian density for the gravitational field only through the tensors that can be constructed from it.
These tensors are the curvature tensor $R^\lambda_{\phantom{\lambda}\rho\mu\nu}=\Gamma^{\,\,\lambda}_{\rho\,\nu,\mu}-\Gamma^{\,\,\lambda}_{\rho\,\mu,\nu}+\Gamma^{\,\,\lambda}_{\sigma\,\mu}\Gamma^{\,\,\sigma}_{\rho\,\nu}-\Gamma^{\,\,\lambda}_{\sigma\,\nu}\Gamma^{\,\,\sigma}_{\rho\,\mu}$, where comma denotes ordinary differentiation, and the torsion tensor $S^\rho_{\phantom{\rho}\mu\nu}=\Gamma^{\,\,\,\,\rho}_{[\mu\,\nu]}$, where square brackets denote antisymmetrization.
We use the notations of \cite{Niko}.
A dynamical affine Lagrangian density should contain derivatives of the connection, so it must contain the curvature tensor.
The simplest algebraic function of derivatives of the connection is a linear function, so it must be proportional to the curvature tensor.
One contraction of the curvature tensor gives the Ricci tensor $R_{\mu\nu}=R^\rho_{\phantom{\rho}\mu\rho\nu}$.
The other contraction gives the antisymmetric, segmental curvature tensor $Q_{\mu\nu}=R^\rho_{\phantom{\rho}\rho\mu\nu}$ \cite{Schou}.
To construct from either tensor a scalar which is linear in derivatives of the connection, we must contract it with a contravariant tensor of rank two which does not contain derivatives of the connection.
Such a tensor must be an algebraic function of the torsion tensor.
The simplest tensors or rank two that are algebraic functions of the torsion tensor are quadratic.
There exist, up to multiplicative constant factors, three such tensors:
\begin{eqnarray}
& & k_{\mu\nu}=S^\rho_{\phantom{\rho}\lambda\mu}S^\lambda_{\phantom{\lambda}\rho\nu}, \label{tens} \\
& & l_{\mu\nu}=S^\rho_{\phantom{\rho}\mu\nu}S_\rho,\,\,\,m_{\mu\nu}=S_\mu S_\nu,
\end{eqnarray}
where $S_\mu=S^\nu_{\phantom{\nu}\mu\nu}$ is the torsion vector.
Their inverses are contravariant tensors and can be used to contract the Ricci tensor or the segmental curvature tensor.

It is always possible, using 3 arbitrary coordinate transformations (in infinitely many ways), to choose the system of reference so that the 3 quantities $l_{0\alpha}$, where $\alpha$ denotes spatial indices, become identically equal to zero.
This choice is analogous to finding a synchronous system of reference, for which $g_{0\alpha}=0$ \cite{LL}.
Accordingly, $\mbox{det}(l_{\mu\nu})=0$ ($l_{00}=0$ by definition) and the inverse of $l_{\mu\nu}$ cannot be constructed.
Similarly, it is always possible, using 1 arbitrary coordinate transformation (in infinitely many ways), to choose the system of reference so that 1 component of $S_\mu$ becomes identically equal to zero.
Accordingly, $\mbox{det}(m_{\mu\nu})=0$ and the inverse of $m_{\mu\nu}$ cannot be constructed.
Consequently, 4 coordinate transformations can be used to impose $\mbox{det}(l_{\mu\nu})=\mbox{det}(m_{\mu\nu})=0$, making the inverses of $l_{\mu\nu}$ and $m_{\mu\nu}$ diverge.
\footnote{
We cannot use 4 coordinate transformations to choose the system of reference so that the 4 quantities $k_{0\mu}$, and thus $\mbox{det}(k_{\mu\nu})$, become identically equal to zero unless $k_{0\mu}=0$ in all systems of reference.
Such a condition, however, would be an artificial constraint on the gravitational field.
}
We therefore use the symmetric, contracted square of the torsion tensor, $k_{\mu\nu}$, in constructing the Lagrangian for the gravitational field.
We define $k^{\mu\nu}$ as a contravariant tensor or rank two which is inverse to $k_{\mu\nu}$:
\begin{equation}
k^{\mu\rho}k_{\nu\rho}=\delta^\mu_\nu.
\end{equation}
This tensor is symmetric, so its contraction with $Q_{\mu\nu}$ vanishes.
The desired scalar is thus $R_{\mu\nu}k^{\mu\nu}$.
Since a Lagrangian density must be a scalar density, we must multiply the above scalar by the square root of the absolute value of the determinant of a covariant tensor of rank two \cite{Ed,Sch,Niko}, which contains the connection and no derivatives of the connection.
Accordingly, we take such a determinant to be
\begin{equation}
k=|\mbox{det}(k_{\mu\nu})|.
\end{equation}
Our affine Lagrangian density of the gravitational field is thus given, up to a multiplicative constant, by
\begin{equation}
\mathfrak{L}_\textrm{g}=R_{\mu\nu}k^{\mu\nu}\sqrt{k}.
\label{Lagr}
\end{equation}
This expression is the simplest dynamical, nondegenerate Lagrangian density that is entirely composed from the affine connection, via its curvature and torsion, and is linear in derivatives of the connection.

To make the Lagrangian density (\ref{Lagr}) meaningful, we must impose on the torsion tensor a condition
\begin{equation}
\mbox{det}(k_{\mu\nu})\neq0.
\label{nonzero}
\end{equation}
If spacetime is continuous, then (\ref{nonzero}) guarantees that the sign of $\mbox{det}(k_{\mu\nu})$ is the same at all points of spacetime.

\section{Affine field equations}

We begin with the gravitational field in the absence of matter.
The variation of the action corresponding to (\ref{Lagr}) is, using $\delta k=-\frac{1}{2}kk_{\mu\nu}\delta k^{\mu\nu}$ \cite{LL},
\begin{equation}
\delta S=\delta\int R_{\mu\nu}k^{\mu\nu}\sqrt{k}d\Omega=\int\delta R_{\mu\nu}k^{\mu\nu}\sqrt{k}d\Omega+\int\Bigl(R_{\mu\nu}-\frac{1}{2}R_{\rho\sigma}k^{\rho\sigma}k_{\mu\nu}\Bigr)\delta k^{\mu\nu}\sqrt{k}d\Omega,
\label{var}
\end{equation}
where $d\Omega$ is an element of four-volume.
Using $\delta R^\lambda_{\phantom{\lambda}\rho\mu\nu}=\delta\Gamma^{\,\,\lambda}_{\rho\,\nu;\mu}-\delta\Gamma^{\,\,\lambda}_{\rho\,\mu;\nu}-2S^\sigma_{\phantom{\sigma}\mu\nu}\delta\Gamma^{\,\,\lambda}_{\rho\,\sigma}$, where semicolon denotes a covariant derivative with respect to the affine connection $\Gamma^{\,\,\rho}_{\mu\,\nu}$, gives $\delta R_{\mu\nu}=\delta\Gamma^{\,\,\rho}_{\mu\,\nu;\rho}-\delta\Gamma^{\,\,\rho}_{\mu\,\rho;\nu}-2S^\sigma_{\phantom{\sigma}\rho\nu}\delta\Gamma^{\,\,\rho}_{\mu\,\sigma}$.
The variation $\delta\Gamma^{\,\,\rho}_{\mu\,\nu}$ is a tensor, and $\int\mathcal{V}^\mu_{\phantom{\mu};\mu}d\Omega=2\int\mathcal{V}^\mu S_\mu d\Omega$ for any contravariant vector density $\mathcal{V}^\mu$ if we omit hypersurface integrals that are set to zero in variational principles \cite{Sch,Niko,Schou}.
The first term on the right of (\ref{var}) therefore becomes
\begin{eqnarray}
& & \int\delta R_{\mu\nu}\,k^{\mu\nu}\sqrt{k}d\Omega=\int(\delta\Gamma^{\,\,\rho}_{\mu\,\nu;\rho}-\delta\Gamma^{\,\,\rho}_{\mu\,\rho;\nu}-2S^\sigma_{\phantom{\sigma}\rho\nu}\delta\Gamma^{\,\,\rho}_{\mu\,\sigma})k^{\mu\nu}\sqrt{k}d\Omega \nonumber \\
& & =\int\bigl[(\delta\Gamma^{\,\,\rho}_{\mu\,\nu}k^{\mu\nu}\sqrt{k})_{;\rho}-(\delta\Gamma^{\,\,\rho}_{\mu\,\rho}k^{\mu\nu}\sqrt{k})_{;\nu}-2S^\sigma_{\phantom{\sigma}\rho\nu}\delta\Gamma^{\,\,\rho}_{\mu\,\sigma}k^{\mu\nu}\sqrt{k}\bigr]d\Omega-\int\delta\Gamma^{\,\,\rho}_{\mu\,\nu}(k^{\mu\nu}\sqrt{k})_{;\rho}d\Omega+\int\delta\Gamma^{\,\,\rho}_{\mu\,\rho}(k^{\mu\nu}\sqrt{k})_{;\nu}d\Omega \nonumber \\
& & =\int\bigl[2S_\rho\delta\Gamma^{\,\,\rho}_{\mu\,\nu}-2S_\nu\delta\Gamma^{\,\,\rho}_{\mu\,\rho}-2S^\sigma_{\phantom{\sigma}\rho\nu}\delta\Gamma^{\,\,\rho}_{\mu\,\sigma}\bigr]k^{\mu\nu}\sqrt{k}d\Omega-\int\delta\Gamma^{\,\,\rho}_{\mu\,\nu}(k^{\mu\nu}\sqrt{k})_{;\rho}d\Omega+\int\delta\Gamma^{\,\,\rho}_{\mu\,\rho}(k^{\mu\nu}\sqrt{k})_{;\nu}d\Omega \nonumber \\
& & =\int\delta\Gamma^{\,\,\rho}_{\mu\,\nu}\bigl[2S_\rho k^{\mu\nu}-2S_\sigma k^{\mu\sigma}\delta^\nu_\rho-2S^\nu_{\phantom{\nu}\rho\sigma}k^{\mu\sigma}\bigr]\sqrt{k}d\Omega-\int\delta\Gamma^{\,\,\rho}_{\mu\,\nu}(k^{\mu\nu}\sqrt{k})_{;\rho}d\Omega+\int\delta\Gamma^{\,\,\rho}_{\mu\,\nu}\delta^\nu_\rho(k^{\mu\sigma}\sqrt{k})_{;\sigma}d\Omega.
\label{first}
\end{eqnarray}
Field equations result from equating to zero the variation (\ref{var}) calculated under a variation $\delta\Gamma^{\,\,\rho}_{\mu\,\nu}$.
The variation $\delta\Gamma^{\,\,\rho}_{\mu\,\nu}$ can be divided into the symmetric part $\delta\Gamma^{\,\,\,\,\rho}_{(\mu\,\nu)}$ and the antisymmetric part $\delta\Gamma^{\,\,\,\,\rho}_{[\mu\,\nu]}=\delta S^\rho_{\phantom{\rho}\mu\nu}$, and we can vary the action under these two parts separately.
We can therefore vary the action under a variation $\delta\Gamma^{\,\,\,\,\rho}_{(\mu\,\nu)}$, substitute the resulting field equation into the action, and then vary the action under a variation $\delta S^\rho_{\phantom{\rho}\mu\nu}$.

We calculate (\ref{var}) with respect to $\delta\Gamma^{\,\,\,\,\rho}_{(\mu\,\nu)}$ and equate it to zero.
Since $\delta k^{\mu\nu}$ in the second term on the right of (\ref{var}) contains only $\delta S^\rho_{\phantom{\rho}\mu\nu}$ and no $\delta\Gamma^{\,\,\,\,\rho}_{(\mu\,\nu)}$, the variation (\ref{var}) with respect to $\delta\Gamma^{\,\,\,\,\rho}_{(\mu\,\nu)}$ is equal to its first term, which is given by the symmetrization of the last line in (\ref{first}) in indices $\mu,\nu$ (denoted by parentheses):
\begin{eqnarray}
& & \delta S_{\Gamma_{()}}=\int\delta\Gamma^{\,\,\,\,\rho}_{(\mu\,\nu)}\bigl[2S_\rho k^{\mu\nu}-2S_\sigma\delta^\nu_\rho k^{\mu\sigma}-2S^\nu_{\phantom{\nu}\rho\sigma}k^{\mu\sigma}\bigr]\sqrt{k}d\Omega-\int\delta\Gamma^{\,\,\,\,\rho}_{(\mu\,\nu)}(k^{\mu\nu}\sqrt{k})_{;\rho}d\Omega+\int\delta\Gamma^{\,\,\,\,\rho}_{(\mu\,\nu)}\delta^\nu_\rho(k^{\mu\sigma}\sqrt{k})_{;\sigma}d\Omega \nonumber \\
& & =\int\delta\Gamma^{\,\,\,\,\rho}_{(\mu\,\nu)}\bigl[(2S_\rho k^{\mu\nu}-2S_\sigma\delta^{(\nu}_\rho k^{\mu)\sigma}-2S^{(\nu}_{\phantom{(\nu}\rho\sigma}k^{\mu)\sigma})\sqrt{k}-(k^{\mu\nu}\sqrt{k})_{;\rho}+\delta^{(\nu}_\rho(k^{\mu)\sigma}\sqrt{k})_{;\sigma}\bigr]d\Omega.
\label{varsym}
\end{eqnarray}
Equating (\ref{varsym}) to zero for arbitrary $\delta\Gamma^{\,\,\,\,\rho}_{(\mu\,\nu)}$ yields
\begin{equation}
(k^{\mu\nu}\sqrt{k})_{;\rho}-\frac{1}{2}\delta^\nu_\rho(k^{\mu\sigma}\sqrt{k})_{;\sigma}-\frac{1}{2}\delta^\mu_\rho(k^{\nu\sigma}\sqrt{k})_{;\sigma}=(2S_\rho k^{\mu\nu}-S_\sigma\delta^\nu_\rho k^{\mu\sigma}-S_\sigma\delta^\mu_\rho k^{\nu\sigma}-S^\nu_{\phantom{\nu}\rho\sigma}k^{\mu\sigma}-S^\mu_{\phantom{\mu}\rho\sigma}k^{\nu\sigma})\sqrt{k}.
\label{consym}
\end{equation}
Contracting (\ref{consym}) with respect to indices $\nu,\rho$ gives $(k^{\mu\sigma}\sqrt{k})_{;\sigma}=\frac{4}{3}S_\nu k^{\mu\nu}\sqrt{k}$, which upon substituting into (\ref{consym}) leads to
\begin{equation}
(k^{\mu\nu}\sqrt{k})_{;\rho}=\Bigl(2S_\rho k^{\mu\nu}-\frac{1}{3}(S_\sigma\delta^\nu_\rho k^{\mu\sigma}+S_\sigma\delta^\mu_\rho k^{\nu\sigma})-S^\nu_{\phantom{\nu}\rho\sigma}k^{\mu\sigma}-S^\mu_{\phantom{\mu}\rho\sigma}k^{\nu\sigma}\Bigr)\sqrt{k}.
\label{fieldcontr}
\end{equation}

Contracting (\ref{fieldcontr}) with $k_{\mu\kappa}k_{\nu\lambda}$, using $(k^{\mu\nu}\sqrt{k})_{;\rho}k_{\mu\kappa}=(\sqrt{k})_{;\rho}\delta^\nu_\kappa-k^{\mu\nu}\sqrt{k}k_{\mu\kappa;\rho}$, gives
\begin{equation}
(\sqrt{k})_{;\rho}k_{\kappa\lambda}-\sqrt{k}k_{\kappa\lambda;\rho}=\Bigl(2S_\rho k_{\kappa\lambda}-\frac{1}{3}(S_\kappa k_{\rho\lambda}+S_\lambda k_{\rho\kappa})-S^\nu_{\phantom{\nu}\rho\kappa}k_{\nu\lambda}-S^\mu_{\phantom{\mu}\rho\lambda}k_{\mu\kappa}\Bigr)\sqrt{k}.
\label{fieldcov}
\end{equation}
Because $\sqrt{k}$ is a scalar density, its covariant derivative is equal to $(\sqrt{k})_{;\rho}=(\sqrt{k})_{,\rho}-\Gamma^{\,\,\sigma}_{\sigma\,\rho}\sqrt{k}$.
The left-hand side of (\ref{fieldcov}) can thus be written as $(\frac{1}{2}k_{\mu\nu,\rho}k^{\mu\nu}-\Gamma^{\,\,\sigma}_{\sigma\,\rho})\sqrt{k}k_{\kappa\lambda}-\sqrt{k}(k_{\kappa\lambda,\rho}-\Gamma^{\,\,\sigma}_{\kappa\,\rho}k_{\sigma\lambda}-\Gamma^{\,\,\sigma}_{\lambda\,\rho}k_{\kappa\sigma})$.
Contracting (\ref{fieldcov}) with $k^{\kappa\lambda}$ gives then $\frac{1}{2}k_{\mu\nu,\rho}k^{\mu\nu}-\Gamma^{\,\,\sigma}_{\sigma\,\rho}=\frac{8}{3}S_\rho$, which upon substituting into (\ref{fieldcov}) leads to
\begin{eqnarray}
k_{\kappa\lambda,\rho}-\Gamma^{\,\,\sigma}_{\kappa\,\rho}k_{\sigma\lambda}-\Gamma^{\,\,\sigma}_{\lambda\,\rho}k_{\kappa\sigma}=k_{\kappa\lambda;\rho}=-Q_{\rho\kappa\lambda},
\label{nonm}
\end{eqnarray}
where
\begin{equation}
Q_{\rho\kappa\lambda}=-\Bigl(\frac{2}{3}S_\rho k_{\kappa\lambda}+\frac{1}{3}S_\kappa k_{\rho\lambda}+\frac{1}{3}S_\lambda k_{\rho\kappa}+S^\sigma_{\phantom{\sigma}\rho\kappa}k_{\sigma\lambda}+S^\sigma_{\phantom{\sigma}\rho\lambda}k_{\sigma\kappa}\Bigr).
\label{nontor}
\end{equation}
Equation (\ref{nonm}) is equivalent to
\begin{equation}
Q_{\kappa\lambda\rho}+Q_{\lambda\rho\kappa}-Q_{\rho\kappa\lambda}=-k_{\lambda\rho,\kappa}-k_{\rho\kappa,\lambda}+k_{\kappa\lambda,\rho}+2\Gamma^{\,\,\,\,\sigma}_{(\kappa\,\lambda)}k_{\rho\sigma}+2S^\sigma_{\phantom{\sigma}\rho\kappa}k_{\lambda\sigma}+2S^\sigma_{\phantom{\sigma}\rho\lambda}k_{\kappa\sigma}.
\label{perm}
\end{equation}
Contracting (\ref{perm}) with $k^{\tau\rho}$ gives
\begin{equation}
\Gamma^{\,\,\,\,\tau}_{(\kappa\,\lambda)}=\{^{\,\,\tau}_{\kappa\,\lambda}\}_k+S^\sigma_{\phantom{\sigma}\kappa\rho}k^{\tau\rho}k_{\lambda\sigma}+S^\sigma_{\phantom{\sigma}\lambda\rho}k^{\tau\rho}k_{\kappa\sigma}-\frac{1}{2}k^{\tau\rho}(Q_{\rho\kappa\lambda}-Q_{\kappa\rho\lambda}-Q_{\lambda\rho\kappa})
\label{sym}
\end{equation}
where
\begin{equation}
\{^{\,\,\rho}_{\mu\,\nu}\}_k=\frac{1}{2}k^{\rho\sigma}(k_{\sigma\nu,\mu}+k_{\sigma\mu,\nu}-k_{\mu\nu,\sigma})
\label{Chrsym}
\end{equation}
are the Christoffel symbols constructed from the tensor $k_{\mu\nu}$.
Using (\ref{nontor}) and (\ref{sym}), we find the affine connection $\Gamma^{\,\,\rho}_{\mu\,\nu}=\Gamma^{\,\,\,\,\rho}_{(\mu\,\nu)}+S^\rho_{\phantom{\rho}\mu\nu}$:
\begin{equation}
\Gamma^{\,\,\rho}_{\mu\,\nu}=\{^{\,\,\rho}_{\mu\,\nu}\}_k+S^\rho_{\phantom{\rho}\mu\nu}-\frac{1}{3}(\delta^\rho_\mu S_\nu+\delta^\rho_\nu S_\mu).
\label{affine}
\end{equation}
This field equation relates the connection to the torsion tensor and its derivatives.

If we define
\begin{equation}
C^\rho_{\phantom{\rho}\mu\nu}=\Gamma^{\,\,\rho}_{\mu\,\nu}-\{^{\,\,\rho}_{\mu\,\nu}\}_k=S^\rho_{\phantom{\rho}\mu\nu}-\frac{1}{3}(\delta^\rho_\mu S_\nu+\delta^\rho_\nu S_\mu),
\label{cont}
\end{equation}
then we can decompose the curvature tensor according to $R^\lambda_{\phantom{\lambda}\rho\mu\nu}=R^{\lambda(k)}_{\phantom{\lambda}\rho\mu\nu}+C^\lambda_{\phantom{\lambda}\rho\nu:\mu}-C^\lambda_{\phantom{\lambda}\rho\mu:\nu}+C^\lambda_{\phantom{\lambda}\sigma\mu}C^\sigma_{\phantom{\sigma}\rho\nu}-C^\lambda_{\phantom{\lambda}\sigma\nu}C^\sigma_{\phantom{\sigma}\rho\mu}$, where $R^{\lambda(k)}_{\phantom{\lambda}\rho\mu\nu}$ is the curvature tensor constructed from the Christoffel symbols (\ref{Chrsym}) instead of the affine connection $\Gamma^{\,\,\rho}_{\mu\,\nu}$, and colon denotes a covariant derivative with respect to these symbols (which also form an affine connection).
The action corresponding to (\ref{Lagr}) is thus
\begin{eqnarray}
& & S=\int R^{(k)}_{\mu\nu}k^{\mu\nu}\sqrt{k}d\Omega+\int(C^\rho_{\phantom{\rho}\mu\nu:\rho}-C^\rho_{\phantom{\rho}\mu\rho:\nu}+C^\lambda_{\phantom{\lambda}\mu\nu}C^\rho_{\phantom{\rho}\lambda\rho}-C^\lambda_{\phantom{\lambda}\mu\rho}C^\rho_{\phantom{\rho}\lambda\nu})k^{\mu\nu}\sqrt{k}d\Omega \nonumber \\
& & =\int R^{(k)}_{\mu\nu}k^{\mu\nu}\sqrt{k}d\Omega+\int(C^\lambda_{\phantom{\lambda}\mu\nu}C^\rho_{\phantom{\rho}\lambda\rho}-C^\lambda_{\phantom{\lambda}\mu\rho}C^\rho_{\phantom{\rho}\lambda\nu})k^{\mu\nu}\sqrt{k}d\Omega+\int\Bigl((C^\rho_{\phantom{\rho}\mu\nu}k^{\mu\nu}\sqrt{k})_{:\rho}-(C^\rho_{\phantom{\rho}\mu\rho}k^{\mu\nu}\sqrt{k})_{:\nu}\Bigr)d\Omega \nonumber \\
& & -\int\Bigl(C^\rho_{\phantom{\rho}\mu\nu}(k^{\mu\nu}\sqrt{k})_{:\rho}-C^\rho_{\phantom{\rho}\mu\rho}(k^{\mu\nu}\sqrt{k})_{:\nu}\Bigr)d\Omega,
\label{decomp}
\end{eqnarray}
where $R^{(k)}_{\mu\nu}=R^{\rho(k)}_{\phantom{\rho}\mu\rho\nu}$.
The relations $\int\mathcal{V}^\mu_{\phantom{\mu}:\mu}d\Omega=0$, which is valid for any contravariant vector density $\mathcal{V}^\mu$ if we omit hypersurface integrals that are set to zero in variational principles \cite{Sch,Niko,Schou}, and $(k^{\mu\nu}\sqrt{k})_{:\rho}=0$ eliminate the last two integrals in (\ref{decomp}).
We also obtain
\begin{eqnarray}
& & C^\lambda_{\phantom{\lambda}\mu\nu}C^\rho_{\phantom{\rho}\lambda\rho}=-\frac{2}{3}l_{\mu\nu}+\frac{4}{9}m_{\mu\nu}, \nonumber \\
& & C^\lambda_{\phantom{\lambda}\mu\rho}C^\rho_{\phantom{\rho}\lambda\nu}=-k_{\mu\nu}-\frac{2}{3}l_{\mu\nu}+\frac{7}{9}m_{\mu\nu}.
\end{eqnarray}
Accordingly, the action (\ref{decomp}) reduces to
\begin{equation}
S=\int\Bigl(R^{(k)}_{\mu\nu}k^{\mu\nu}+4-\frac{1}{3}m_{\mu\nu}k^{\mu\nu}\Bigr)\sqrt{k}d\Omega.
\label{act}
\end{equation}
The last term in the parentheses on the right of (\ref{act}) can be set to zero by one arbitrary coordinate transformation.
The first two terms have the form of the Einstein-Hilbert Lagrangian for the gravitational field with a cosmological constant in the general theory of relativity, if we identify, up to a multiplicative constant factor, the tensor $k_{\mu\nu}$ with the metric tensor $g_{\mu\nu}$.

If we define a nondimensional tensor
\begin{equation}
g_{\mu\nu}=\frac{2}{\Lambda}k_{\mu\nu},
\label{dim}
\end{equation}
where a constant $\Lambda$ has the dimension of inverse square length, then the Christoffel symbols constructed from $g_{\mu\nu}$ satisfy $\{^{\,\,\rho}_{\mu\,\nu}\}_g=\{^{\,\,\rho}_{\mu\,\nu}\}_k$.
The corresponding curvature and Ricci tensors satisfy respectively $R^{\lambda(g)}_{\phantom{\lambda}\rho\mu\nu}=R^{\lambda(k)}_{\phantom{\lambda}\rho\mu\nu}$ and $R^{(g)}_{\mu\nu}=R^{(k)}_{\mu\nu}$.
The first two terms on the right of (\ref{act}) become the Einstein-Hilbert action with the cosmological constant $\Lambda$:
\begin{equation}
S\propto\int(R^{(g)}_{\mu\nu}g^{\mu\nu}+2\Lambda)\sqrt{g}d\Omega,
\end{equation}
where $g=|\mbox{det}(g_{\mu\nu})|$.
The tensor $g_{\mu\nu}$ therefore corresponds to the metric tensor in general relativity.
The definition of the metric tensor (\ref{dim}) gives the square of the interval, $ds^2=g_{\mu\nu}dx^\mu dx^\nu$, in terms of the torsion tensor:
\begin{equation}
ds^2=\frac{2}{\Lambda}S^\rho_{\phantom{\rho}\lambda\mu}dx^\mu S^\lambda_{\phantom{\lambda}\rho\nu}dx^\nu=D^\rho_{\phantom{\rho}\lambda}D^\lambda_{\phantom{\lambda}\rho}=\mbox{tr}(D^2),
\end{equation}
where the matrix $D$ is defined as
\begin{equation}
D^\rho_{\phantom{\rho}\lambda}=\Bigl(\frac{2}{\Lambda}\Bigr)^{1/2}S^\rho_{\phantom{\rho}\lambda\mu}dx^\mu.
\end{equation}

To make the definition (\ref{dim}) meaningful, we must use the physical signature requirement for $g_{\mu\nu}$ by choosing the sign of $\mbox{det}(k_{\mu\nu})$, which is the same at all points of spacetime, to be negative.
Taking into account only those configurations with
\begin{equation}
\mbox{det}(k_{\mu\nu})<0
\label{Lorentz}
\end{equation}
will guarantee that the metric tensor has the Lorentzian signature and can therefore be reduced at a given point to the Minkowski tensor of flat spacetime.

The relation (\ref{dim}) gives the cosmological constant the physical meaning: this constant provides a length scale of the affine connection.
Together with $c$ and $G$, $\Lambda$ thus provides a natural system of units (length, time, mass) in classical gravitational physics.
In quantum physics, a natural (Planck) system of units is defined by $\hbar$, $c$ and $G$.
The length scale associated with the observed cosmological constant, $(\Lambda)^{-1/2}\sim10^{26}\mbox{m}$, is 61 orders of magnitude larger than the Planck length.
The origin of such an enormous value is yet to be found.

In the presence of matter, we must add to the Lagrangian density for the gravitational field (\ref{Lagr}) the Lagrangian density for the matter, $\mathfrak{L}_\textrm{m}$.
The action for the gravitational field and matter is then
\begin{equation}
S=\int(\mathfrak{L}_\textrm{g}+\alpha\mathfrak{L}_\textrm{m})d\Omega,
\label{matter}
\end{equation}
where $\alpha$ is a constant that sets the unit of mass (and is related to Newton's gravitational constant).
If we consider matter fields that do not couple to the affine connection through a covariant derivative, then they can depend on the connection only through the torsion tensor.
The variation of the action (\ref{matter}) with respect to $\Gamma^{\,\,\,\,\rho}_{(\mu\,\nu)}$ is thus not affected by such fields, resulting in (\ref{varsym}).
Accordingly, the field equation (\ref{affine}) remains valid.
However, if we include matter fields that couple to the affine connection through a covariant derivative, such as Dirac spinors, then (\ref{affine}) will be modified by terms which depend on these fields.
To describe fermionic matter, we must define beforehand the metric tensor using (\ref{dim}) with the torsion tensor satisfying (\ref{Lorentz}).
Consequently, a Lorentz subgroup can be defined and spinors can be constructed.

In the presence of matter fields that depend on the connection only through the torsion tensor, the action (\ref{act}) becomes
\begin{equation}
S=\int\Bigl(R^{(k)}_{\mu\nu}k^{\mu\nu}+4-\frac{1}{3}m_{\mu\nu}k^{\mu\nu}\Bigr)\sqrt{k}d\Omega+\alpha\int\mathfrak{L}_\textrm{m}d\Omega.
\label{action}
\end{equation}
We consider the Lagrangian density for matter $\mathfrak{L}_\textrm{m}$ that depends on the torsion tensor only through $k_{\mu\nu}$.
Varying the action (\ref{action}) with respect to $\delta S^\rho_{\phantom{\rho}\mu\nu}$ and equating it to zero yields the field equation for the torsion tensor.
Following (\ref{var}), we have
\begin{eqnarray}
& & \delta S=\int\delta R^{(k)}_{\mu\nu}k^{\mu\nu}\sqrt{k}d\Omega+\int\Bigl(R^{(k)}_{\mu\nu}-\frac{1}{2}R^{(k)}_{\rho\sigma}k^{\rho\sigma}k_{\mu\nu}-2k_{\mu\nu}-\frac{1}{3}S_\mu S_\nu+\frac{1}{6}S_\rho S_\lambda k^{\rho\lambda}k_{\mu\nu}\Bigr)\delta k^{\mu\nu}\sqrt{k}d\Omega \nonumber \\
& & -\frac{2}{3}\int\delta^{[\nu}_\rho S_\lambda k^{\mu]\lambda}\delta S^\rho_{\phantom{\rho}\mu\nu}\sqrt{k}d\Omega+\alpha\int\frac{\delta\mathfrak{L}_\textrm{m}}{\delta k^{\mu\nu}}\delta k^{\mu\nu}d\Omega.
\label{varia}
\end{eqnarray}
The first term on the right of (\ref{varia}) gives
\begin{eqnarray}
& & \int\delta R^{(k)}_{\mu\nu}k^{\mu\nu}\sqrt{k}d\Omega=\int(\delta\{^{\,\,\rho}_{\mu\,\nu}\}_{k:\rho}-\delta\{^{\,\,\rho}_{\mu\,\rho}\}_{k:\nu})k^{\mu\nu}\sqrt{k}d\Omega=\int\Bigl((\delta\{^{\,\,\rho}_{\mu\,\nu}\}_{k}k^{\mu\nu}\sqrt{k})_{:\rho}-(\delta\{^{\,\,\rho}_{\mu\,\rho}\}_{k}k^{\mu\nu}\sqrt{k})_{:\nu}\Bigr)d\Omega \nonumber \\
& & -\int\Bigl(\delta\{^{\,\,\rho}_{\mu\,\nu}\}_{k}(k^{\mu\nu}\sqrt{k})_{:\rho}-\delta\{^{\,\,\rho}_{\mu\,\rho}\}_{k}(k^{\mu\nu}\sqrt{k})_{:\nu}\Bigr)d\Omega=0,
\end{eqnarray}
similarly to the last two integrals in (\ref{decomp}).
Substituting an identity $\delta k^{\mu\nu}=-k^{\mu\rho}k^{\nu\sigma}\delta k_{\rho\sigma}=-2S^\tau_{\phantom{\tau}\kappa\sigma}k^{\mu\rho}k^{\nu\sigma}\delta S^\kappa_{\phantom{\kappa}\tau\rho}$ into (\ref{varia}) leads to
\begin{eqnarray}
& & \delta S=-2\int\Bigl(R^{(k)}_{\kappa\lambda}-\frac{1}{2}R^{(k)}_{\pi\tau}k^{\pi\tau}k_{\kappa\lambda}-2k_{\kappa\lambda}+\frac{\alpha}{\sqrt{k}}\frac{\delta\mathfrak{L}_\textrm{m}}{\delta k^{\kappa\lambda}}-\frac{1}{3}S_\kappa S_\lambda+\frac{1}{6}S_\pi S_\tau k^{\pi\tau}k_{\kappa\lambda}\Bigr)S^{[\mu}_{\phantom{[\mu}\rho\sigma}k^{\nu]\kappa}k^{\lambda\sigma}\delta S^\rho_{\phantom{\rho}\mu\nu}\sqrt{k}d\Omega \nonumber \\
& & -\frac{2}{3}\int\delta^{[\nu}_\rho S_\lambda k^{\mu]\lambda}\delta S^\rho_{\phantom{\rho}\mu\nu}\sqrt{k}d\Omega.
\label{varanti}
\end{eqnarray}
Equating (\ref{varanti}) to zero for arbitrary $\delta S^\rho_{\phantom{\rho}\mu\nu}$ yields
\begin{eqnarray}
& & \Bigl(R^{(k)}_{\kappa\lambda}-\frac{1}{2}R^{(k)}_{\pi\tau}k^{\pi\tau}k_{\kappa\lambda}-2k_{\kappa\lambda}+\frac{\alpha}{\sqrt{k}}\frac{\delta\mathfrak{L}_\textrm{m}}{\delta k^{\kappa\lambda}}-\frac{1}{3}S_\kappa S_\lambda+\frac{1}{6}S_\pi S_\tau k^{\pi\tau}k_{\kappa\lambda}\Bigr)S^{[\mu}_{\phantom{[\mu}\rho\sigma}k^{\nu]\kappa}k^{\lambda\sigma}+\frac{1}{3}\delta^{[\nu}_\rho S_\lambda k^{\mu]\lambda}=0,
\end{eqnarray}
which is satisfied if
\begin{eqnarray}
& & S_\mu=0, \nonumber \\
& & R^{(k)}_{\mu\nu}-\frac{1}{2}R^{(k)}_{\rho\sigma}k^{\rho\sigma}k_{\mu\nu}=2k_{\mu\nu}-\frac{\alpha}{\sqrt{k}}\frac{\delta\mathfrak{L}_\textrm{m}}{\delta k^{\mu\nu}}.
\label{Einst}
\end{eqnarray}
Substituting (\ref{dim}) into (\ref{Einst}) and using the definition of the metric energy-momentum tensor of matter $T_{\mu\nu}$, $\delta\mathfrak{L}_\textrm{m}=\frac{1}{2}T_{\mu\nu}\sqrt{g}\delta g^{\mu\nu}$, gives
\begin{equation}
R^{(g)}_{\mu\nu}-\frac{1}{2}R^{(g)}_{\rho\sigma}g^{\rho\sigma}g_{\mu\nu}=\Lambda g_{\mu\nu}-\frac{\alpha}{\Lambda}\frac{2}{\sqrt{g}}\frac{\delta\mathfrak{L}_\textrm{m}}{\delta g^{\mu\nu}}=\Lambda g_{\mu\nu}-\frac{\alpha}{\Lambda}T_{\mu\nu}.
\end{equation}
This equation reproduces the Einstein equations of general relativity with the cosmological constant, if we regard
\begin{equation}
G=-\frac{\alpha c^4}{8\pi\Lambda}
\end{equation}
as the gravitational constant.

The field equations (\ref{Einst}) comprise 4 equations for $S_\mu$ and 10 equations for $k_{\mu\nu}$, giving 14 equations for the 24 components of the torsion tensor.
If the matter fields depend on the torsion tensor only through $k_{\mu\nu}$, then 10 components of $S^\rho_{\phantom{\rho}\mu\nu}$ are undetermined.
At least 10 components of the torsion tensor propagate because the equations for $k_{\mu\nu}$ are second-order differential equations.
To compare, MacDowell and Mansouri have synthesized the 24 spin connection variables and 16 tetrad variables of the first-order formalism into a generalized connection containing 40 variables \cite{MM}.

\section{Summary}

The affine theory of gravitation presented here provides a natural construction of the metric tensor solely from the affine connection (through the torsion tensor), as its quadratic function.
This construction is reverse to that in GR, where the connection is composed from the metric and the Lagrangian density has the simplest dynamical form.
Taking the simplest dynamical Lagrangian density that is entirely composed from the connection and is linear in its derivatives leads to the field equations which are equivalent to the Einstein equations in vacuum with a cosmological constant.
Including matter fields that couple to torsion only through the tensor $k_{\mu\nu}$ reproduces the Einstein equations with the energy-momentum tensor and a cosmological constant.
Since the cosmological constant appears naturally as a consequence of our affine formulation, the observed dark energy is simply an intrinsic property of spacetime.
The observed accelerating expansion of the Universe may therefore be the strongest evidence for torsion, favoring the affine formulation.

The Eddington-Schr\"{o}dinger purely affine gravity, in which the metric tensor is defined from the total Lagrangian for the gravitational field and matter, also generates a cosmological constant in the field equations.
Such a tensor is not, however, a purely geometrical object constructed solely from the affine connection; its definition involves the matter fields.
In our theory, on the contrary, the metric tensor is an algebraic function of the connection only; it depends on the matter fields through the dynamical field equations.
Our construction of the metric tensor as a quadratic function of the torsion tensor arises naturally from the field equations, which become the Einstein equations of GR if $k_{\mu\nu}$ is interpreted as $g_{\mu\nu}$ (up to a constant multiplicative factor which becomes the cosmological constant).

Including matter fields that couple to the affine connection not only through $k_{\mu\nu}$, such as Dirac spinors, should result in the field equations that are either equivalent to the Einstein-Cartan field equations of the ECSK theory of gravity (with a cosmological constant) or very similar.
We expect that the our construction of the metric tensor as a quadratic function of the torsion tensor will reproduce the construction of the metric as a function of Dirac spinors \cite{spin}.
The affine theory of gravitation in the presence of Dirac fields will be studied in a subsequent paper.

If the metric tensor is composed of the torsion tensor, then it is not a fundamental quantity describing the gravitational field.
Accordingly, the concept of the graviton as an elementary particle associated with the metric and mediating the gravitational force becomes unphysical.
It is also possible that the concept of the metric is entirely classical.
The metric tensor could be proportional to the expectation value $\langle\hat{k}_{\mu\nu}\rangle$ of a quantum operator $\hat{k}_{\mu\nu}=\hat{S}^\rho_{\phantom{\rho}\lambda\mu}\hat{S}^\lambda_{\phantom{\lambda}\rho\nu}$.

The affine formulation of gravity is conceptually similar to gauge theories of other fundamental forces. 
In the presence of a gravitational field, we generalize an ordinary derivative into a coordinate-covariant derivative by introducing the affine connection, while in the presence of the electromagnetic field we generalize it into a $U(1)$-covariant derivative by introducing the electromagnetic potential.
The connection is the dynamical variable in the affine theory of gravitation, like the electromagnetic potential in classical electrodynamics.

\end{document}